\documentclass[lettersize,journal]{IEEEtran}
\usepackage{amsmath,amsfonts}
\usepackage{algpseudocode}

\usepackage{algorithm}
\usepackage{array}
\usepackage[caption=false,font=normalsize,labelfont=sf,textfont=sf]{subfig}
\usepackage{textcomp}
\usepackage{stfloats}
\usepackage{url}
\usepackage{verbatim}
\usepackage{graphicx}
\usepackage{cite}

\usepackage[nolist,nohyperlinks]{acronym}
\usepackage{todonotes}

\begin{document}

\begin{acronym}
	\acro{rfsoc}[RFSoC]{Radio-Frequency System-on-Chip}
	\acro{soc}[SoC]{System-on-Chip}
	\acro{mkid}[MKID]{Microwave Kinetic Inductance Detector}
	\acro{mmc}[MMC]{Magnetic Microcalorimeter}
	\acro{squid}[SQUID]{Superconducting Quantum Interference Device}
	\acro{pl}[PL]{Programmable Logic}
	\acro{ps}[PS]{Processing System}
	\acro{dac}[DAC]{Digital-Analog Converter}
	\acro{adc}[ADC]{Analog-Digital Converter}
	\acro{vna}[VNA]{Vector Network Analyzer}
	\acro{iir}[IIR]{Infinite Impulse Response}
	\acro{qic}[QIC]{Quantum Interface Controller}
	\acro{sdr}[SDR]{Software-Defined Radio}
	\acro{rfdc}[RFdc]{RF data converter}
	\acro{cirque}[cirque]{Communication interface to readout electronics for quantum experiments}
	\acro{fir}[FIR]{Finite Impulse Response}
	\acro{hil}[HIL]{Hardware-in-the-Loop}
	\acro{tes}[TES]{Transition-Edge Sensor}
	\acro{umux}[$\mu$MUX]{Microwave SQUID Multiplexer}
	\acro{lfsr}[LFSR]{Linear Feedback Shift Register}
    \acro{sm}[SM]{Standard Model}
    \acro{tdd}[TDD]{Test-Driven Development}
    \acro{nco}[NCO]{Numerically Controlled Oscillator}
    \acro{rmse}[RMSE]{Root-Mean-Squared Error}
    \acro{mae}[MAE]{Mean Absolute Error}
    \acro{dsp}[DSP]{Digital Signal Processing}
    \acro{pmf}[PMF]{Probability Mass Function}
    \acro{lut}[LUT]{Look-up Table}
    \acro{fdm}[FDM]{Frequency-Division Multiplexing}
\end{acronym}

\title{CryoDE: a Digital Cryogenic Detector Emulator for Microwave SQUID Multiplexed Systems}

\author{
T.~Muscheid, 
D.~Crovo, 
R.~Gartmann, 
E.~Gerlein, 
O.~Sander, 
S.~Kempf, 
L.~E.~Ardila-Perez  

\thanks{
Manuscript received September 26, 2025; revised August 16, 2021.

L.~E.~Ardila-Perez, R.~Gartmann, T.~Muscheid (e-mail: timo.muscheid@kit.edu), and O.~Sander are with the Institute for Data Processing and Electronics, Karlsruhe Institute of Technology, 76344 Eggenstein-Leopoldshafen, Germany.

D.~Crovo is with the Department of Electronics, Pontificia Universidad Javeriana, Bogotá, Colombia and the Institute for Data Processing and Electronics, Karlsruhe Institute of Technology, 76344 Eggenstein-Leopoldshafen, Germany.

E.~Gerlein is with the Department of Electronics, Pontificia Universidad Javeriana, Bogotá, Colombia

S.~Kempf is with the Institute of Micro- and Nanoelectronic Systems, Karlsruhe Institute of Technology, 76187 Karlsruhe, Germany and the Institute for Data Processing and Electronics, Karlsruhe Institute of Technology, 76344 Eggenstein-Leopoldshafen, Germany.
}

}

\markboth{Journal of \LaTeX\ Class Files,~Vol.~14, No.~8, August~2021}%
{Shell \MakeLowercase{\textit{et al.}}: A Sample Article Using IEEEtran.cls for IEEE Journals}

\IEEEpubid{0000--0000/00\$00.00~\copyright~2021 IEEE}

\maketitle

\begin{abstract}
Simultaneous readout of large-scale cryogenic detector arrays relies on multiplexing schemes such as the \ac{fdm} with microwave \ac{squid} multiplexers and highly customized readout electronics.
In traditional detector systems, where mixed-signal ASICs are used in detector front-ends and typically provide a digital interface, \ac{hil} testing can be readily implemented by reusing the existing digital logic of the front-end for emulation purposes. Such straightforward emulation is not possible for \ac{fdm} low-temperature detectors, where the sensor signal is encoded in a high-frequency microwave carrier via a two-stage modulation scheme depending on the cryogenic resonators and the SQUID response.
To address this challenge, we present CryoDE, a digital cryogenic-detector emulator for microwave \ac{squid} multiplexed detector systems. CryoDE generates the encoded detector signals, including realistic pulse responses, enabling full \ac{hil} testing of the room-temperature DAQ system without requiring the cryogenic hardware. This resource-efficient FPGA-based detector twin integrates seamlessly into existing DAQ systems and allows experiment-specific adjustment of detector-signal parameters.
We describe the internal architecture and capabilities of CryoDE within our custom \ac{hil} framework and demonstrate its use in evaluating the performance of real-time signal processing firmware optimized for different microwave SQUID multiplexed cryogenic-detector experiments.
\end{abstract}


\section{Introduction}

\IEEEPARstart{T}{he pursuit} of deeper understanding in fundamental physics has consistently driven major technological breakthroughs, particularly in the development of experimental setups for probing complex phenomena. 
For example, high-energy experiments in particle physics helped stablish the \ac{sm} of Particle Physics in the 1970s, a foundational framework that explains subatomic particles and the fundamental forces they interact with, and later inspired the construction of the Large Hadron Collider (LHC), which culminated in the discovery of the Higgs boson---the long-predicted but previously undetected cornerstone of the \ac{sm}.

Despite the \ac{sm}’s success, unresolved questions---such as the neutrino mass---indicate the existence of physics beyond the Standard Model (BSM). Modern experiments searching for BSM physics increasingly rely on cryogenic detectors to meet their scientific goals, employing several thousand individual detectors to increase both the count rate and the energy resolution. 
An efficient technique for operating large-scale \ac{mmc} or \ac{tes} arrays is microwave SQUID multiplexing, which integrates high-bandwidth Radio-Frequency-SQUIDs (RF-SQUIDs) and modulates the signals across GHz-range resonators \cite{irwin_2004}. For instance, experiments aiming to measure the neutrino mass, perform X-ray imaging spectroscopy or measure the polarization of the cosmic microwave background are planning to utilize this measurement technique \cite{Holmes, ECHo, TOMCAT, Simons}.

Readout of these \ac{umux} systems poses significant challenges for the room-temperature DAQ electronics. The DAQ is responsible for resonator stimulation, frequency-domain demultiplexing, demodulation, and pulse triggering, while simultanously avoiding any degradation of the detector sensitivity through additional noise. A thorough evaluation and characterization of these systems require the complete measurement setup, including an operational cryostat with the \ac{umux} and low-temperature detectors. However, since measurement time is strictly limited and valuable, and because both DAQ systems and detectors are often designed and developed in parallel and not always available at the same time, a full evaluation of DAQ systems is often not feasible. Instead, the DAQ is typically operated at room temperature in loopback mode, directly demultiplexing the generated frequency comb without the modulation of the RF-SQUID. This approach allows only a partial characterization of the system, omitting the testing of algorithms for flux-ramp demodulation and pulse triggering.

\IEEEpubidadjcol
In this work, we present CryoDE, a digital emulator for \ac{umux} detection systems. This module emulates both the detectors and the RF-SQUIDs, effectively reproducing the entire cryogenic setup. As the emulator is implemented as a VHDL module, it can be conveniently integrated into the existing firmware of a readout electronics system, allowing full evaluation of the system in a \ac{hil} setup. Additionally, the emulator can be used for \ac{tdd}. The readout systems are usually tailored to the experiment-specific requirements. By adjusting the emulator parameters accordingly, the system can be verified already at the early stages of firmware conceptualization and implementation. For instance, CryoDE enables the iterative development of custom trigger algorithms specifically designed for particular use cases.
In Section~\ref{sec_sys_reqs}, we define the specific requirements of the emulator and then provide an in-depth explanation of its architecture and implementation details in Section~\ref{sec_impl}. In Section~\ref{sec_eval}, we demonstrate the usability of the emulator by integrating it into the firmware of the readout system developed for the ECHo-100k experiment.

\section{System Requirements}
\label{sec_sys_reqs}
The development of CryoDE was driven by the need for a convenient method to evaluate current and future readout electronics for \ac{umux} systems. Our readout systems are implemented using the \ac{sdr} approach, with a Zynq Ultrascale+ MPSoC as the main element. In the first step, CryoDE is integrated directly into the readout firmware on the SoC’s programmable logic (PL). In the future, we plan to develop a standalone emulation system that can interface with the readout electronics via SMA connectors. In particular, we need the emulator to generate a signal that can be processed by our readout firmware, which consist of a frequency-demultiplexing stage followed by flux-ramp demodulation and pulse triggering. The emulated signal has to mimic a frequency comb where each tone is modulated by the SQUID response to a flux-ramp signal, with the detector signals encoded in the phase of the SQUID. For realistic detector signals, randomized pulses are generated with a distribution that follows actual decay rates.

To meet certain scientific goals, each experiment employs different detector and \ac{umux} designs, thereby imposing application-specific requirements on the readout electronics. CryoDE is intended to be a generic module that can be integrated into the readout electronics of various experiments. Its parameters can be configured either at build-time or dynamically at run-time. Fundamental design choices, such as the number of detectors in the array or the detector response after energy deposition, are defined at early stages of the conceptualization; hence, the emulator is configured statically to these numbers during firmware creation. To thorougly characterize the frequency demultiplexing, the resonance frequencies of the multiplexer must be adjustable at run-time. Similarly, evaluating the flux-ramp demodulation with various SQUID response frequencies and amplitudes is crucial. For analyzing the performance of the trigger algorithms, adjustable count rates and amplitudes of the emulated detector signals are beneficial.

To simplify integration of the emulator into the existing firmware, the top-level VHDL module is equipped with an AXI4-Stream interface. Since all other modules in our library also use this protocol for input and output data streams, connecting CryoDE in the block diagram is straightforward. To account for the effects of the analog front-end into the characterization, the emulator should be placed in the digital domain upstream of the DAC output. 
For user control, such as changing configuration parameters, an AXI-lite interface was chosen, combined with our custom ``servicehub'' gRPC server implemented in the Linux processing system \cite{Karcher_2021}. CryoDE is implemented in a modular way, where each instance emulates a single channel of the \ac{umux}, consisting of one resonator and one RF-SQUID coupled to it. For emulation of a full multiplexer with multiple channels, the emulator module has to be instantiated multiple times in the firmware, followed by a subsequent module that sums the output signals. This approach ensures full flexibility and broad usability of the emulator across different use cases.
Based on these choices, CryoDE was conceptualized and implemented. The details of the implementation are described in Section~\ref{sec_impl}.

\section{System Architecture}
\label{sec_impl}

\begin{figure}[htbp]
	\centering
	\includegraphics[width=\columnwidth]{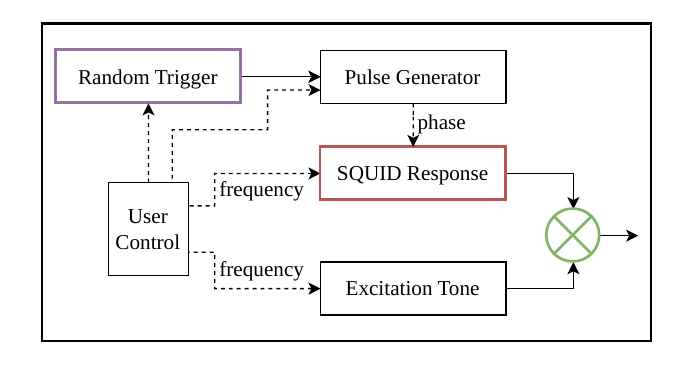}
	\caption{Top-level architecture of CryoDE for a single \ac{umux} channel. The three blocks in the top row emulate the detector signal, which is used as the phase-modulating input for the SQUID response. The colored borders around certain blocks are explained in more detail in Subsections~\ref{subsec_trigger} and \ref{subsec_squid_response}}
	\label{fig_detector_emulator}
\end{figure}

The required features of the emulator can be divided into several parts, allowing a modular implementation approach. Figure~\ref{fig_detector_emulator} shows the top-level architecture of CryoDE. The detector signal is emulated by a pulse generator that is activated using a random trigger at the input. The predefined pulse shape is subsequently scaled in amplitude to generate different pulse heights according to user-defined parameters. The SQUID response is generated by an NCO configured with the user-defined frequency and amplitude, using the emulated detector signal as phase input. Finally, this phase-modulated SQUID signal is amplitude-modulated onto the carrier tone, resulting in the output signal of a single channel of the \ac{umux}. All modules generate two complex samples per clock cycle at the output, achieving a maximum baseband frequency of 1\,GHz when operated with a clock frequency of 500\,MHz. In the following, the individual submodules of the emulator system are described in more detail.

\subsection{Random Trigger}
\label{subsec_trigger}

\begin{figure}[htbp]
	\centering
	\includegraphics[width=0.9\columnwidth]{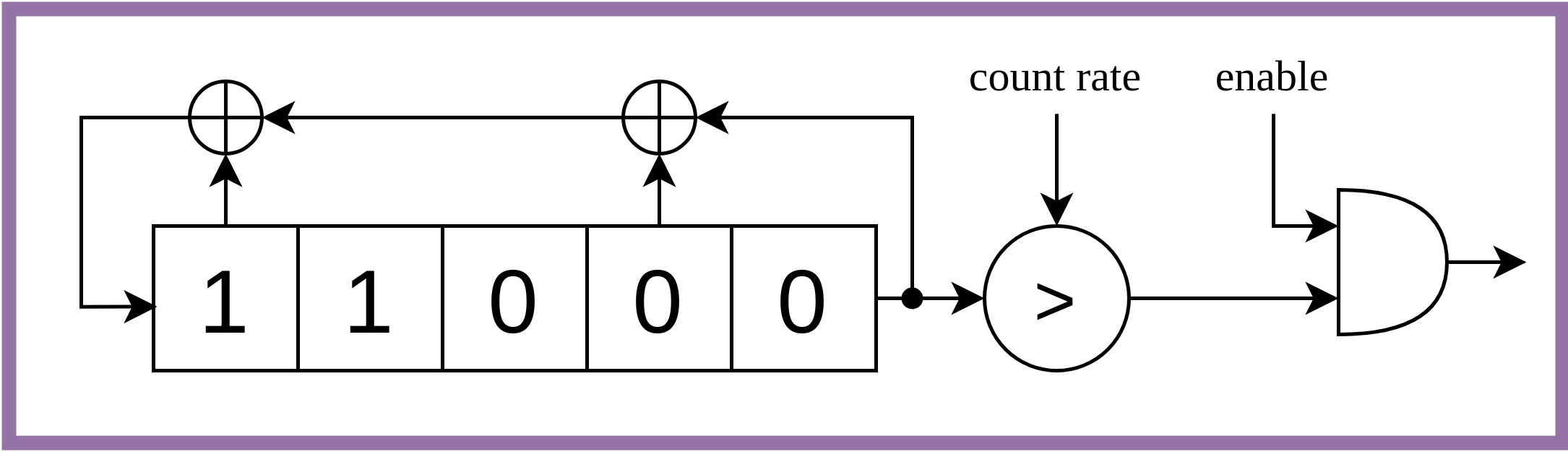}
	\caption{Model of the SQUID response. Due to flux-ramp modulation, the SQUIDs periodically shift the resonance frequency. Using a static carrier tone at frequency $f_{exc}$, this frequency shift results in a change of amplitude. For simplicity, this amplitude modulation is approximated by a sine wave.}
	\label{fig_detector_emulator_trigger}
\end{figure}

The random trigger module generates a 1-bit signal to trigger the generation of a single pulse, thereby emulating the decay of radioactive sources. This decay process can be modeled as a Poisson process, in which decay events occur randomly over time and can, for small time intervals, be approximated by a Bernoulli trial \cite{veiga_2016, turner_2012}. A 96-bit \ac{lfsr} pseudo-random number generator is implemented to produce 32-bit uniformly distributed random numbers. The output of the \ac{lfsr} is used to perform a Bernoulli trial in each clock cycle by comparing the 32-bit number with a threshold defined by a user-configurable parameter that sets the count rate. If the \ac{lfsr} output is below this threshold, the trigger signal is set to ’0’, otherwise to ’1’. Additionally, the output can be suppressed by deactivating the enable signal. A diagram of the trigger with a simplified version of the \ac{lfsr} is depicted in Figure~\ref{fig_detector_emulator_trigger}. The module allows the user to define the count rate---the average number of decay events expected per second---corresponding to decay rates ranging from 0.5 to 35184\,Bq.

\subsection{Radioactive Pulse Generator}

The shape of the detector response to energy deposition strongly depends on the design of the detector itself. To maintain flexibility and support multiple applications, we chose a \ac{lut}-based approach to synthesize the exponentially decaying signals. The waveform \(f(t)\) is precomputed in software with non-signed 16-bit resolution and 12-bit address width, and is stored in a file that initializes a Block RAM (BRAM) at build time. The module handles the amplitude and time scaling of the waveform, adjusting each sample to align with the clock rate and the expected timing of the synthesized signal. When a trigger signal is received, the module starts reading the BRAM by decrementing the \ac{lut} index. If a subsequent trigger is received and the current signal has reached 80\% of its decay time constant value $\tau$, the \ac{lut} index is reset; otherwise, the current read cycle completes uninterrupted. The final output is a non-signed 16-bit signal reflecting the expected behavior of an energy deposition in the \ac{mmc} detector.

The pulse generation process relies on sequentially reading values from the \ac{lut}. Since the sampling time step \(\Delta t\) for \(f(t)\) differs from the target clock period \(T_c\), a synchronization mechanism is necessary to ensure that the output aligns with each \(T_c\). This alignment guarantees that the correct values are generated at the intended times, namely when \(T_c \approx t_n\). One common solution involves linear interpolation between consecutive \ac{lut} samples to approximate intermediate amplitude values. However, this approach adds extra multiplication and computation per output sample, and CryoDE aims to minimize real-time processing requirements. Hence, this emulator adopts a simplified method that uses discrete amplitude levels for each time step, as described in Algorithm~\ref{alg:pulse_generation}, where:
\begin{itemize}
    \item \(T_c\) is the operating clock period.
    \item \(\Delta t\) is the sampling time step used for \(f(t)\) when populating the \ac{lut}.
    \item \(t_c\) is a time accumulator incremented by \(T_c\) on each clock cycle.
    \item \(t_n\) is a time accumulator incremented by \(\Delta t\), i.e.\ \( t_n = n \cdot \Delta t \).
    \item \(n\) is the index for accessing the \ac{lut} entries.
\end{itemize}

\begin{algorithm}[tbh]
\caption{Pulse Generation from \ac{lut}}
\label{alg:pulse_generation}
\begin{algorithmic}
\Require {trigger, $T_c$, $\Delta t$}
\Ensure{pulse\_output}
\State $state \gets IDLE $ 
\State $pulse\_output \gets 0 $ 
\State $n, \ t_c , \ t_n \gets 0$
\While{$clock \And enable$}
    \If{state = IDLE}
        \State $pulse\_output \gets 0$
        \If{trigger = 1}
            \State $state \gets RUNNING$
        \EndIf
    \ElsIf{state = RUNNING}
        \State $t_c \gets t_c + T_c$            
        \If{$n = N $}
            \State $state \gets IDLE $ 
            \State $pulse\_output \gets 0 $ 
            \State $n, \ t_c , \ t_n \gets 0$
        \EndIf
        \If{$t_c \leq t_n$}
            \State $pulse\_output \gets LUT[n]$
        \Else
            \State $pulse\_output \gets LUT[n]$
            \State $t_n \gets t_n + \Delta t$            
            \State $n \gets n + 1$

        \EndIf
    \EndIf
\EndWhile
\end{algorithmic}
\end{algorithm}

The algorithm operates in two states. In the \texttt{IDLE} state, all variables and the output are reset to zero until the trigger signal is asserted, which initiates the pulse generation process. The state machine then transitions to the 
\texttt{RUNNING} state, where the following steps are executed on each clock cycle:

\begin{itemize}
    \item The current emulator time, \(t_c\), is incremented by \(T_c\), tracking the elapsed time according to the system clock period.
    \item If \(t_c \leq t_n\), the output remains assigned to the \ac{lut} sample at index \(n\). No time accumulators or \ac{lut} indices are incremented because \(t_c\) has not yet caught up to \(t_n\).
    \item If \(t_c > t_n\), a full \(\Delta t\) interval has elapsed in the emulator's timing, so \(n\) is incremented to move to the next \ac{lut} sample, and \(t_n\) is updated accordingly.
    \item When \(n = N\), a complete pulse has been generated. The algorithm transitions back to the \texttt{IDLE} state, and the output pulse and time accumulators are reset to zero.
\end{itemize}

This design methodology ensures that the output pulses accurately reproduce the intended exponential decay profile by directly correlating the emulator’s time accumulators with the \ac{lut} indices.

\subsection{SQUID response generation}
\label{subsec_squid_response}

\begin{figure}[htbp]
	\centering
	\includegraphics[width=0.9\columnwidth]{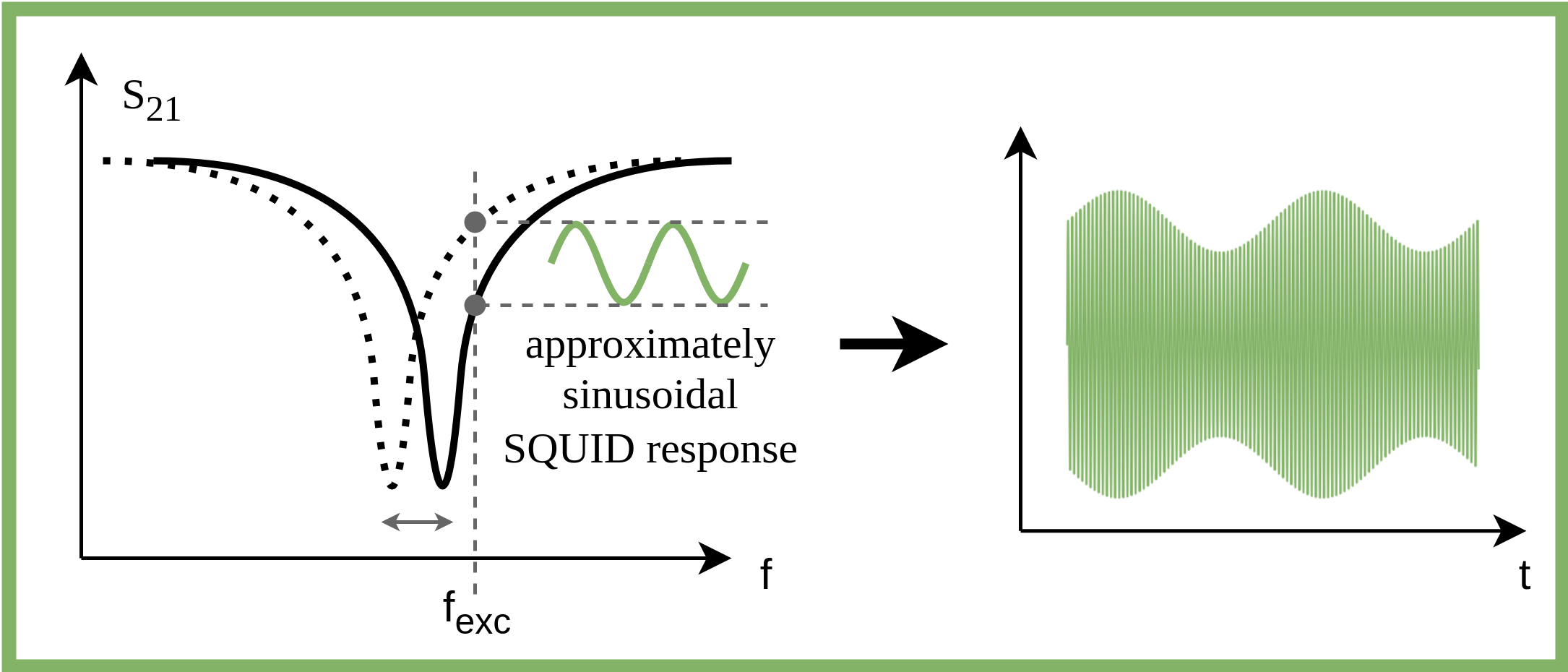}
	\caption{Model of the SQUID response. Due to flux-ramp modulation, the SQUIDs periodically shift their resonance frequency. Using a static carrier tone at frequency $f_{exc}$, this frequency shift appears as a change in amplitude. For simplicity, this amplitude modulation is modeled as a sine wave.}
	\label{fig_detector_emulator_model}
\end{figure}

On a \ac{umux}, the detectors are inductively coupled to an RF-SQUID which is itself coupled to the resonator. Due to the intrinsic non-linearity of the SQUIDs, they need to be linearized using flux-ramp modulation. The SQUID response to this modulation can be modeled as a sine wave, as shown in Figure~\ref{fig_detector_emulator_model}. This sine wave is generated using an \ac{nco} with user-defined frequency and amplitude as inputs. By varying the step size and the scaling factor accordingly, the \ac{nco} iterates through a memory containing precomputed sine-wave samples to generate the desired frequency. 

\begin{figure}[htbp]
	\centering
	\includegraphics[width=0.9\columnwidth]{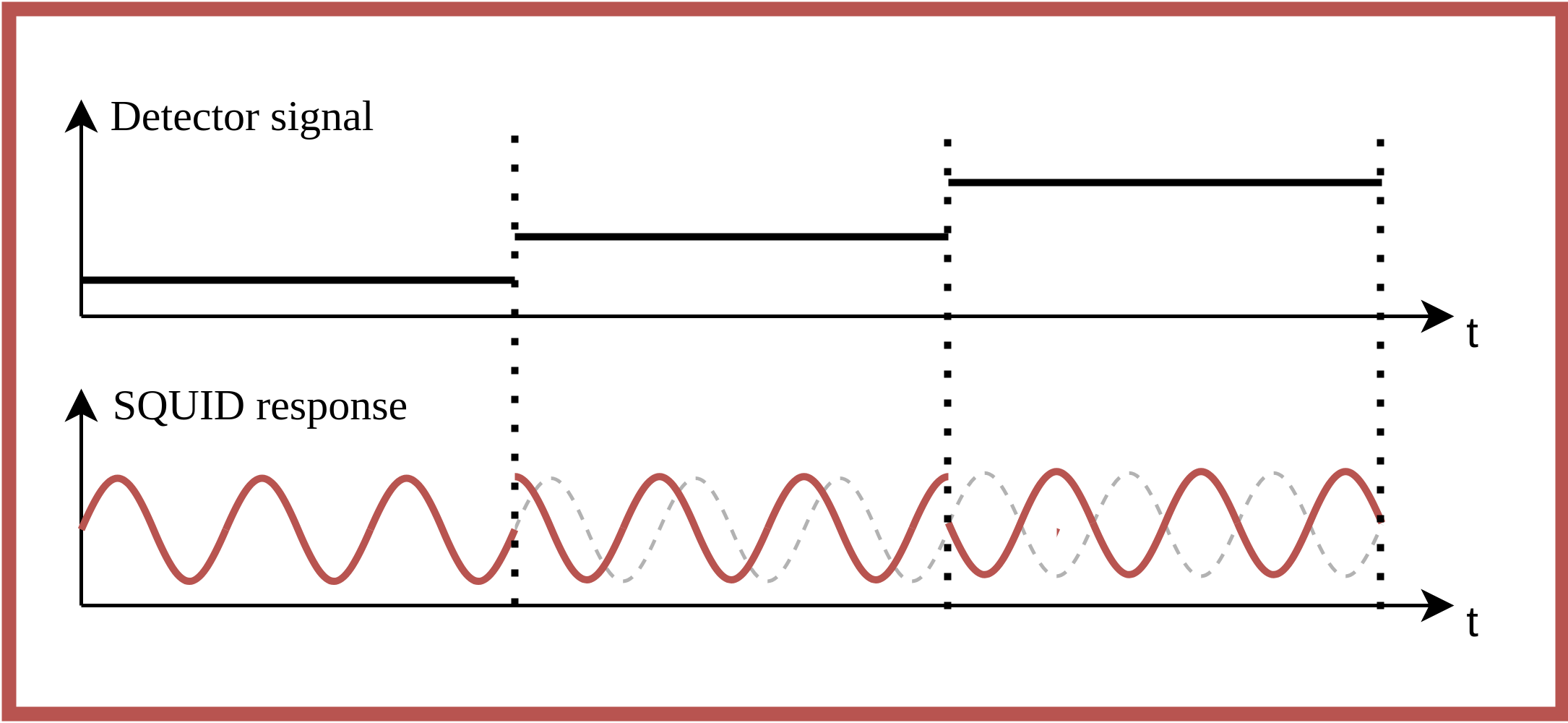}
	\caption{Phase encoding of the detector signal in the SQUID response. For successful demodulation of the signal, the frequency of the SQUID response must be significantly higher than the bandwidth of the detector signal, ensuring that the phase-encoded signal remains effectively constant within at least one period of the SQUID response.}
	\label{fig_detector_emulator_phase}
\end{figure}

The emulated detector signal containing the pulses is used as the phase input of the \ac{nco}, resulting in a phase-encoded signal, as shown in Figure~\ref{fig_detector_emulator_phase}. In our \ac{nco} implementation, this phase signal is interpreted as an offset to the read-pointer address, and any change in the input leads to a jump of the read address.

\subsection{Carrier signal generation}

Currently, CryoDE does not use the full frequency comb as the input signal to emulate a resonator. Instead, the specific carrier tone used for resonator readout is generated internally using an \ac{nco}. This approach enables simple frequency shifting of the signal to characterize the readout electronics across various frequency bands. The generated carrier signal is amplitude-modulated by the SQUID response signal using a \ac{dsp} slice to multiply the two signals.


\section{Evaluation}
\label{sec_eval}

For evaluation, CryoDE was tested using the ECHo-100k experiment as a reference application. The ECHo-100k experiment employs \acp{mmc} to measure the spectrum of the electron-capture process of \textsuperscript{163}Holmium, aiming to determine the upper limit of the electron-neutrino mass. The room-temperature readout electronics \cite{Muscheid_2024} enable operation of a \ac{umux} with 400 resonators operating in the frequency band between 4 to 8\,GHz. The activity of the detectors is designed to be 10\,Bq each. We integrated four instances of CryoDE into the firmware, configured them to match the requirements, and analyzed the output. 
As described in Section~\ref{subsec_trigger}, decay processes follow a Poisson distribution with a \ac{pmf} for $\lambda$ expected events:

\begin{equation}
	\begin{aligned}
		PMF(k) = \frac{\lambda ^{k} * e^{- \lambda}}{k!}
	\end{aligned}
	\label{eq_pulse_shape}
\end{equation}

\begin{figure}
	\centering
    \subfloat{\includegraphics[height=0.45\columnwidth]{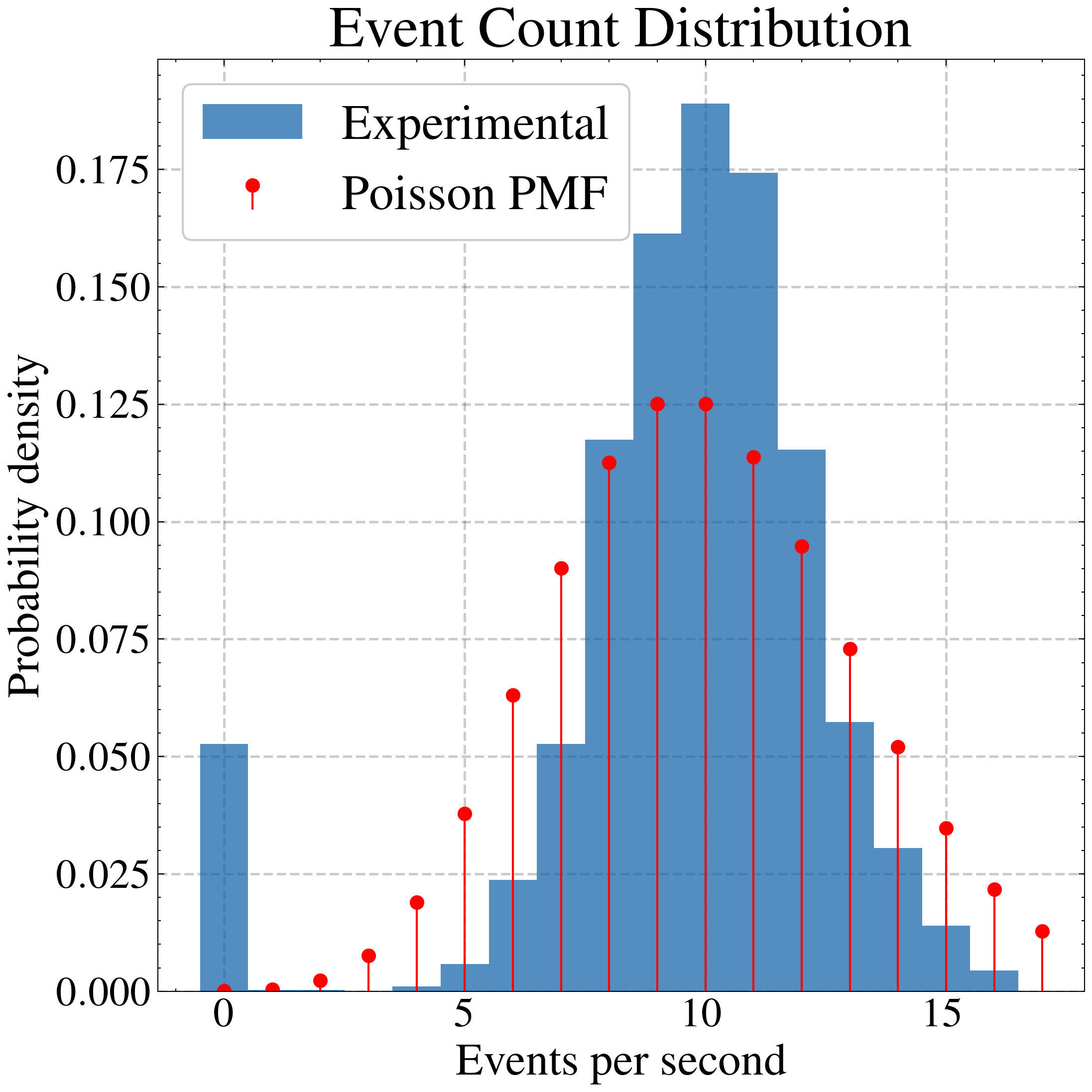}}
    \hfill
	\subfloat{\includegraphics[height=0.45\columnwidth]{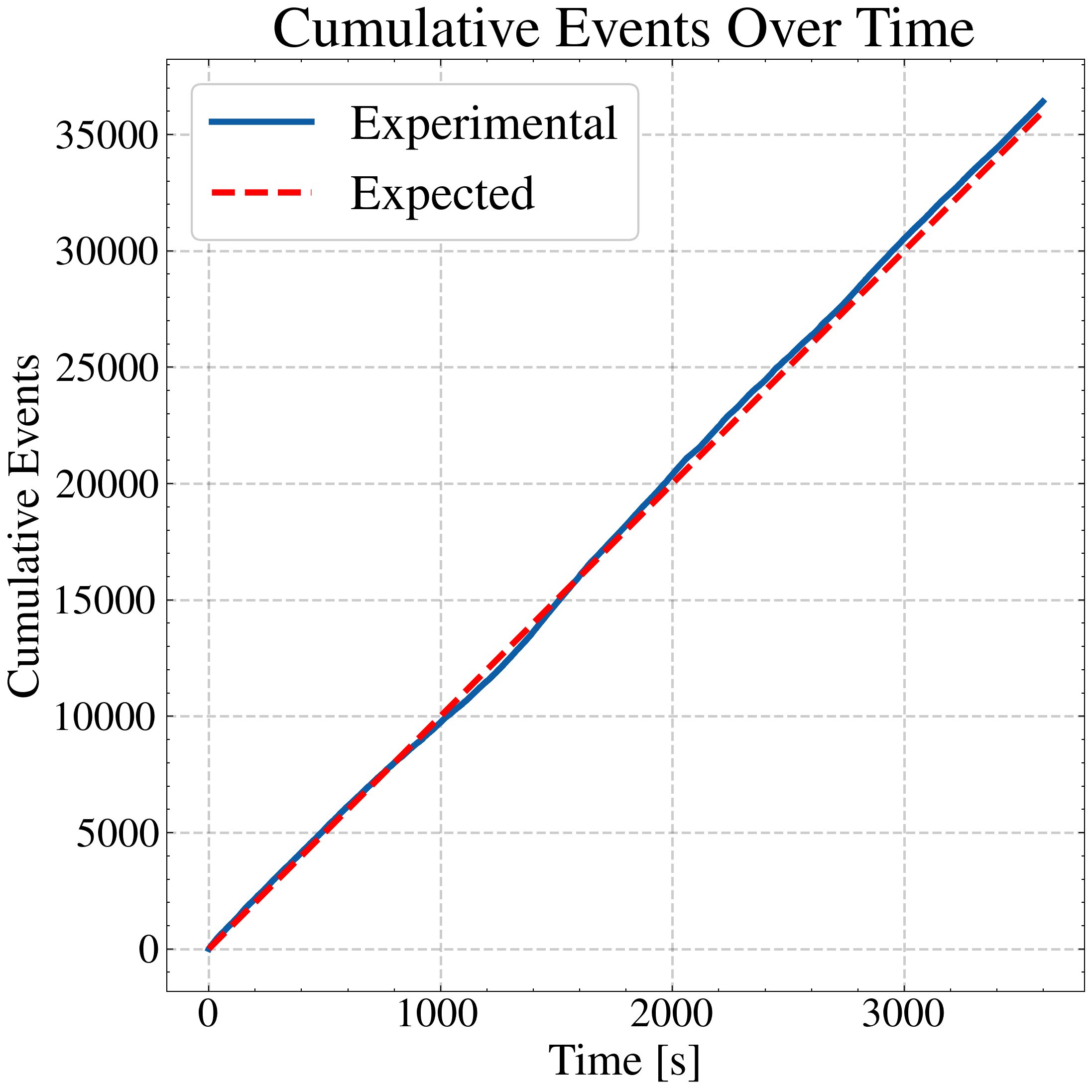}}
	\caption{Characterization of the emulator performance with a count rate set to 10\,Bq. On the left, the distribution of the actual event count is compared to the \ac{pmf} of the theoretical Poisson distribution. The right plot illustrates the long-term behavior of the event rate compared to the expected cumulative events.}
	\label{fig_count_rate_evaluation}
\end{figure}

Figure~\ref{fig_count_rate_evaluation} shows that the experimental event distribution roughly follows the Poisson distribution with an outlier at 0 events. This deviation is due to our model using a \ac{lfsr} as the random-number generator, which is not truly random. However, the average count rate matches the specifications and remains stable in the long term.

\begin{figure}[htbp]
	\centering
	\includegraphics[width=\columnwidth]{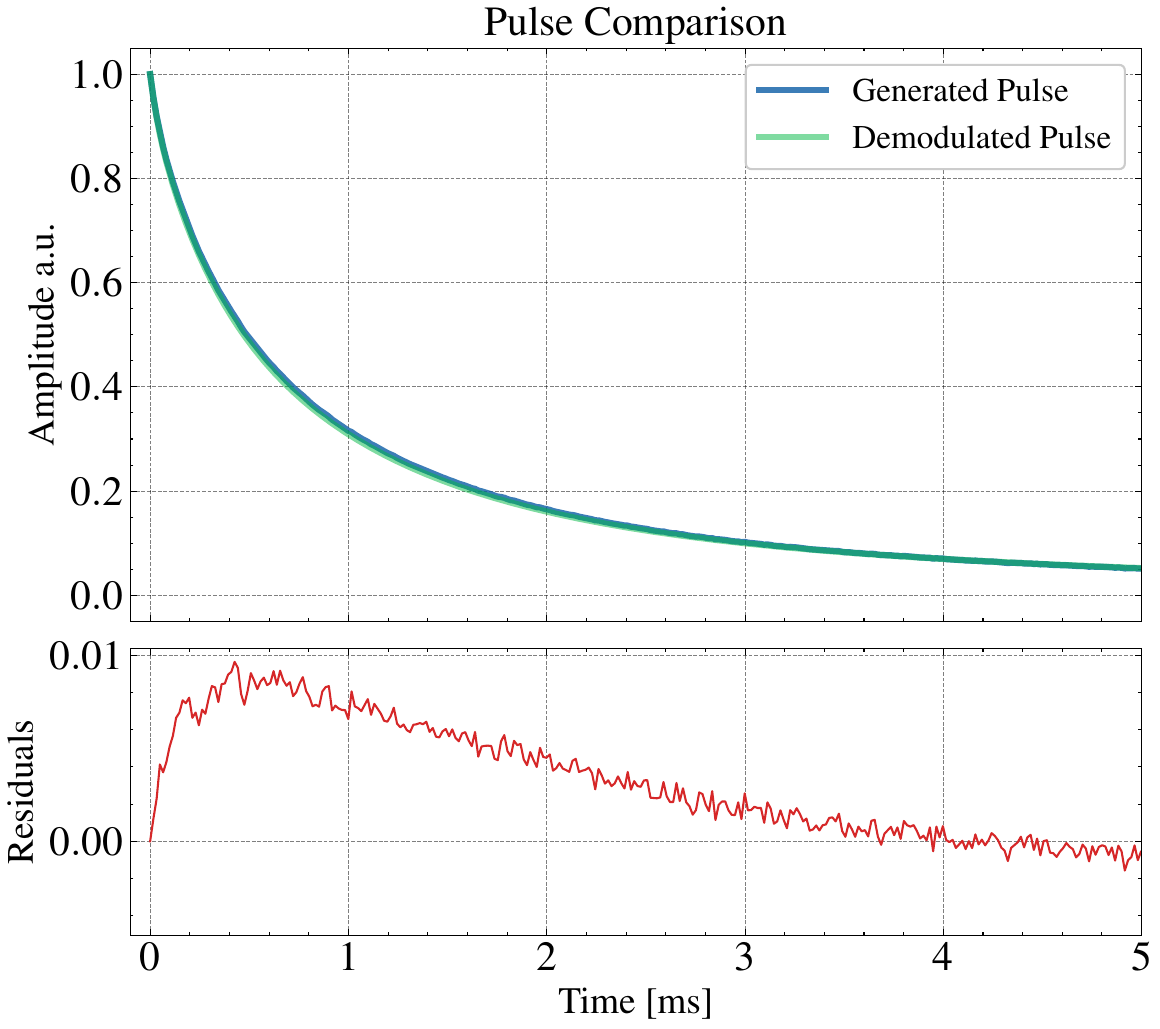}
	\caption{Comparison between the generated pulse shape and the triggered signal after frequency demultiplexing and flux-ramp demodulation.}
	\label{fig_pulse_comparison}
\end{figure}

The typical signal shape of an \ac{mmc} after the energy of a particle is deposited in the absorber is modeled by the following equation \cite{schaefer_2012}:

\begin{equation}
	\begin{aligned}
		\Delta\Phi(t) = A_{0} * \left[ \frac{\tau_{1}}{\tau_{1} - \tau_{r}}\left( e^{-t/\tau_{1}} - e^{-t/\tau_{r}}\right)  \right.\\
       \left. - \frac{\tau_{0}}{\tau_{0} - \tau_{r}}\left( e^{-t/\tau_{0}} - e^{-t/\tau_{r}} \right) \right]
	\end{aligned}
	\label{eq_pulse_shape}
\end{equation}

with $A_{0}$ defining the pulse amplitude, $\tau_{0}, \tau_{1}$ representing the time constants for the signal rise and decay, respectively, and $\tau_{r}$ denoting the cutoff frequency of the readout electronics. In reality, the signal decay is not a single exponential function but rather a sum of four exponential functions \cite{schaefer_2012}. However, the exact shape of the pulse is not critical here, as the purpose of CryoDE is to assess the accuracy of reconstruction of the readout system. The residuals in Figure~\ref{fig_pulse_comparison} provide insight into this measure, indicating that the demodulated signal is in good agreement with the generated pulse, with a deviation less than 1\,\%.

A more detailed analysis of the reconstruction accuracy is conducted using the \ac{rmse} and the \ac{mae}. These metrics provide a general understanding of the average deviation between the generated pulse and the \ac{mmc} signal. The \ac{rmse} is more sensitive to larger errors due to its quadratic formulation, while the \ac{mae} represents an average of the absolute error and is less sensitive to significant deviations.
The results presented in Table~\ref{tab:error_metrics} show that the modulation and the demodulation process on the platform do not noticeably degrade the signal.

\begin{table}[ht]
\centering
\caption{Error Metrics for Comparing Generated Pulses}
\label{tab:error_metrics}
\def\arraystretch{3}
\begin{tabular}{|ccc|}
\hline
\textbf{Metric} & \textbf{Formula} & \textbf{Value} \\
\hline
\ac{rmse} & $\sqrt{ \frac{1}{N} \sum_{j=1}^{N} \left( f(j)-\text{data}[j] \right)^2 }$ & 0.0032 \\
\ac{mae} & $\frac{1}{N} \sum_{j=1}^{N} \left| f(j) -\text{data}[j] \right|$ & 0.0026 \\
R² & $1 - \frac{\sum_{j=1}^{N} (f(j) - \text{data}[j])^2}{\sum_{j=1}^{N} (f(j) - \bar{f})^2}$ & 0.9995 \\
\hline
\end{tabular}
\end{table} 

One key goal during the development of CryoDE was to maintain resource efficiency to enable integration of the system into existing firmware. For the measurements described here, the emulator was integrated into the Zynq Ultrascale+ RFSoC ZU49DR. The resources consumed by one instance of the emulator are listed in Table~\ref{tab:fpga_utilization} and are compared with both the total resources available on the device and the resources consumed by the full-scale ECHo-100k firmware image.

\begin{table}[ht]
\centering
\caption{Resource utilization}
\label{tab:fpga_utilization}
\begin{tabular}{|l|cccc|}
\hline
\textbf{Module} & \textbf{LUT} & \textbf{FF} & \textbf{Memory} & \textbf{DSP} \\
\hline
Available & 425280 & 850560 & 60.5 Mb & 4272 \\
ECHo-100k firmware & 407460 & 654630 & 50 Mb & 1375 \\
CryoDE & 872 & 1383 & 810 kb & 10\\
\hline
\end{tabular}
\end{table} 

This resource-utilization profile shows that the implementation of CryoDE effectively balances performance requirements with resource constraints, while preserving sufficient headroom for future enhancements.

\section{Conclusion}

The emulator for cryogenic detectors read out via \ac{umux} presented here enables fully self-contained firmware verification and calibration. The FPGA-based module is customizable to the requirements of specific applications, with many of the parameters adaptable at runtime to allow measurements under different conditions. The randomized event rate approximately follows a Poisson distribution and remains stable over long periods. This makes CryoDE well-suited for evaluating the performance of trigger algorithms or even for developing new algorithms using the \ac{tdd} approach. Integrating CryoDE into the room-temperature readout electronics for the ECHo-100k experiment enabled an analysis of the performance of the firmware. The results showed that the reconstructed pulse after frequency demultiplexing and flux-ramp demodulation is consistent with the pulse generated by CryoDE, confirming the accuracy of our DAQ system.


\end{document}